\documentstyle[12pt]{article}

\begin{document}

\baselineskip=5mm

\begin{center}

{Number of states with fixed angular momentum for identical fermions and bosons }

\vspace{0.3in}

{ Y. M. Zhao$^{a,b}$ and  
  A. Arima$^{c}$}

\vspace{0.3in}

{ $^a$ Department of Physics, Saitama University, Saitama-shi, Saitama 338 Japan \\
$^b$ Department of Physics,  Southeast University, Nanjing 210018 China \\
$^c$ The House of Councilors, 2-1-1 Nagatacho, 
Chiyodaku, Tokyo 100-8962, Japan }

\date{\today}

\end{center}

We present in this paper  
empirical formulas for the number of angular momentum $I$ states  
for three and four identical fermions or bosons. 
In the cases with  large $I$ we prove that the number of 
states  with  the same     ${\cal M}$ and $n$ 
but different $J$ is identical if 
$I \ge (n-2)J - \frac{1}{2} (n-1)(n-2)$  for fermions and  
$I \ge (n-2)J$ for bosons, and that 
the number of states is also identical 
for the same ${\cal M}$ but different $n$ and $J$ if
${\cal M} \le $min($n$, $2J+1 - n$)  for fermions and
for ${\cal M} \le $min($n$, $2J$)  for bosons. 
Here  ${\cal M} =I_{max}-I$, 
 $n$ is the particle number, and $J$ refers to 
  the angular momentum of a single-particle orbit
for  fermions, or the spin $L$  carried by  bosons.

\vspace{0.3in}

        
\newpage

The enumeration of the number  of total angular momentum $I$ 
states is a very common practice in nuclear structure theory.  
One usually obtains this number 
by subtracting the   combinatorial  number  of the states with 
a total  angular momentum 
projection $M=I+1$ from that for  $M=I$.  
The    combinatorial numbers of different $M$'s seem  
to be  irregular, and such a enumeration procedure 
would be prohibitively tedious without a computer for 
a very large single-$j$ shell.   In textbooks \cite{Ring} 
the numbers of states of a few nucleons  in a single-$j$ 
($j$ is a half integer) shell are usually
tabulated for sake of convenience.  
There was only one algebraic formulas available  
for $I=0$ states with  $4$ fermions \cite{Ginocchio}.
It would be very interesting to have  
more general algebraic formulas, if possible. 
In another context, it was noticed \cite{Zhao} that 
the angular momentum 0 ground state probability 
of 4 fermions in a single-$j$ shell in the presence of random interactions 
 has a synchronous staggering with an increase of the number of $I=0$ states
 by one. It has not been clarified yet  whether there is a deep relation
between these two quantities or this
synchronism is just a coincidence. 
This ``coincidence" also  motivated the present study of number of states. 
 
In this paper we  give some
simple  formulas for the  
number of states,
denoted as $D(I)_j$, $D(I)_l$, and $D(I)_L$
for  fermions in a single-$j$ shell, fermions in a  single-$l$ shell, 
and  bosons with  spin $L$, respectively, 
where $j$ is a half integer, and
$l$ and $L$ are integers. We are able to construct empirical 
formulas for all $I$ states of  three and four identical  particles,  and 
some formulas for a few lowest $I$ states of five particles. 
 For short we  use
$D(I)_J$ to refer to all these three cases, where 
$J$ can be $j$, $l$, or $L$. 
Although  these formulas are obtained empirically, 
we have confirmed them  by computer for 
$j\le 999/2$, $l\le 500$ or   $L\le 500$,
and for many cases with  much larger $J$'s which were taken randomly. 
One therefore can use them ``safely"  in practice. 
  We shall also show  that the number of  
states is identical for the same $n$ and ${\cal M}$ (defined as $I_{max}-I$) 
but different $J$ for states of fermions with 
$I \ge  (n-2) J - \frac{1}{2} (n-1)(n-2)$
or for states of bosons with $I \ge (n-2)L$,
 and that the number of states  with  
different $n$ and $J$ is also identical if 
${\cal M}  \le $min($n$, $2J+1 - n$)  for fermions and
 ${\cal M} \le $min($n$, $2J$)  for bosons. 
In the Appendix we shall list 
 a few formulas for the cases of five particles.  

For $n=3$ and $I \le J$, we empirically obtain 
\begin{eqnarray}
&& D(I)_j = \left[ \frac{2I+3}{6}  \right] ~ ;  \nonumber \\ 
&& D(I)_l = \left[ \frac{I}{3}  \right] +
\frac{1}{2} \left( 1 - (-)^{I+l} \right) ~ ;  \nonumber \\ 
&& D(I)_L = \left[ \frac{I}{3}  \right] + 
\frac{1}{2} \left( 1+ (-)^{I+L} \right) ~, 
\end{eqnarray}
where $\left [ ~~ \right]$ means to take the largest  integer not exceeding
the value inside.

For  fermions with $n=3$ and $I \ge J-1$    or bosons with 
$n=3$ and $I \ge J$, the $D(I)_J$'s can be empirically given in a 
unified form:  
\begin{equation}
D(I)_J = \left[ \frac{I_{max}-I }{6}  \right] + \delta_I,  
\end{equation}
where
\begin{eqnarray}
&& \delta_I   =  \left\{
\begin{array}{ll}
0      &  {\rm if}~ (I_{max}-I)\%6 = 1   \\
1       &  {\rm otherwise} ~.  \nonumber 
\end{array}  \right.
\label{n3_1}
\end{eqnarray}
In this paper  $a\%b$ means to take the remainder of $a/b$, where
$a$ is a non-negative integer and $b$ is a natural number. For
examples, $7 \%3 =1$, $27 \%10 =7$. 

According to Eq. (1),  
 $D(\frac{1}{2})_j=0$, $D(0)_l=1$ (or 0) if $l$ is odd (or even),
 and $D(0)_L=1$ (or 0) if $l$ is  even (or odd). 
It is noted that there are overlaps of $I$'s covered by  Eqs. (1) and (2),
and that one may use either of them to obtain the number of states for 
these $I$'s.

The cases  of four particles are more complicated. However, a 
regular staggering of $D(I)_J$ can be easily 
noticed if $I \le 2J$.  
We define $\eta_I^J = D(I)_{J+1} - D(I)_{J}$,
and $\Delta(I) = D(I)_{J} - D(I+3)_J$ (even $I$). It is
found that 
$\eta_I^J$ changes periodically as $J$ changes by 3 
for $I \le 2J$.  Table I shows that the  ($\eta_I^J$, $\eta_I^{J+1}$,
$\eta_I^{J+2}$) of  four particles 
is  actually a  ``trinary representation"  of 
a natural number $(I/2+1)$, where $I$ is an even number.
Although the origin of this regularity is not known,  
one may make use of this to  
 construct the formulas of  $D(I)_J$ for $n=4$ and $I \le 2J$. 
The cases of $I\ge 2J$ will be addressed  later in this paper.

For $n=4$ and $I \le 2J$ with $I$ being even, 
we empirically obtain 
\begin{equation}
D(I)_J =  \left[ \frac{ {\cal L} - I/2 }{3} \right] \times
(\frac{I}{2} + 1) + C(I)  m
- \delta + {\cal D} (I), 
\label{n4_f}
\end{equation}
where $m = ({\cal L} - I/2) \% 3$.
For four fermions in a single-$j$ shell,
${\cal L}=j-\frac{1}{2}$,
and the  coefficients in Eq. (\ref{n4_f}) are given by 
\begin{eqnarray}
&&  C(I)  = \left[ \frac{I}{6} \right] + 1 ~ , 
\nonumber \\
&& \delta   =  \left\{
\begin{array}{ll}
 \delta_{m2}      &  {\rm if}~ I\% 6 =0 ~     \\
 0      &  {\rm otherwise} ~   
\end{array}  \right. ~  , \nonumber \\
&&  {\cal D} (I) = 3 K (K-1) + K +
 \left( 1 + {\cal K} \right) K +
 \delta_{{\cal K}, 4} +  \delta_{{\cal K}, 5} ~,    \nonumber 
\end{eqnarray}
where $K = \left[  \frac{I+4}{12} \right]$,
and ${\cal K} = \frac{ (I+4) \% 12}{2}$. 
For four fermions in a single-$l$ shell,
${\cal L}=l$, 
and the  coefficients in Eq. (\ref{n4_f}) are given by 
\begin{eqnarray}
&&  C(I)  = \left[ \frac{I+4}{6} \right]   ~ , 
\nonumber \\
&& \delta   =  \left\{
\begin{array}{ll}
 \delta_{m2}      &  {\rm if}~ (I+4)\% 6 =0 ~     \\
 0      &  {\rm otherwise} ~  
\end{array}  \right. ~ , \nonumber \\
&&
 {\cal D} (I) = 3 K (K-1) + K +
 \left( 1 + {\cal K} \right) K +
 \delta_{{\cal K}, 4} +  \delta_{{\cal K}, 5} ~,    \nonumber 
\end{eqnarray}
where $K = \left[  \frac{I+2}{12} \right]$,
and ${\cal K} = \frac{ (I+2) \% 12}{2}$. 
For 4 bosons with  spin $L$,
${\cal L}=L$, 
and  the coefficients in Eq. (\ref{n4_f}) are given by 
\begin{eqnarray}
&&  C(I)_J = \left[ \frac{I+4}{6} \right]  ~ , 
\nonumber \\
&& \delta   =  \left\{
\begin{array}{ll}
 \delta_{m2}      &  {\rm if}~ (I+4)\% 6 =0 ~     \\
 0      &  {\rm otherwise} ~  
\end{array}  \right. ~ , \nonumber \\
&&
 {\cal D}_L(I) = 3 K (K-1) + K +
 \left( 1 + {\cal K} \right) K +
 \delta_{{\cal K}, 4} +  \delta_{{\cal K}, 5} ~,  \nonumber 
\end{eqnarray}
where $K = \left[  \frac{I+8}{12} \right]$,
and ${\cal K} = \frac{ (I+8) \% 12}{2}$.

For  $n=4$ and $I \le 2J$ with $I$ being odd, 
we introduce  $I = I_0 + 3$, and obtain
\begin{eqnarray}
&& D(I)_j =  D_j(I_0) - \left[ \frac{I}{4} \right] - 1 ~ ,  \nonumber \\
&& D(I)_l =  D_l(I_0) - \left[ \frac{I+2}{4} \right] ~ ,  \nonumber \\
&& D(I)_L =  D_L(I_0) - \left[ \frac{I}{4} \right] - 1 ~  .
\label{n4_f2}
\end{eqnarray}
One easily sees that the $D(1)_J$'s are always zero for $n=4$.

For  $n=4$ and $I \ge 2J$, 
we define
\begin{eqnarray}
&  {\rm for ~~ even} & I  : ~ ~ I = I_{max} - 2m ~,  \nonumber \\
&  {\rm for ~~ odd } & I  : ~ ~ I = I_{max}-3 - 2m   ~. \nonumber  
\end{eqnarray}
We let $K= \left[ \frac{m}{6} \right]$, ${\cal K}= m \% 6$, and 
 obtain 
\begin{equation}
D(I)_J = 3K (K+1) -K + (K+1)({\cal K} +1) + \delta_{{\cal K} 0}  -1 
\label{n4_f3}
\end{equation}
for fermions with $I \ge   2J - \frac{1}{2} (n-1) (n-2) = 2J -3$
and for bosons with $I \ge 2L = I_{max} -2L$.

 It is noted that for fermions $D(I)_J$ of $I=(2J - 3,
2J -2, 2J -1, $ and $2J)$  can be obtained either
by Eqs.  (\ref{n4_f}) and (\ref{n4_f2}) or Eq. (\ref{n4_f3}). 
The formulas of $D(I)_J$ for $n=4$ present 
an even-odd staggering of the number of states: the number of states
with even number of $I$ is  not smaller and mostly larger than those
of their odd $I$ neighbors.  
A similarity between the formulas for four fermions (in both 
half-integer $j$ orbit and integer $l$ orbit)
and bosons is also easily noticed.

The situation of  $n=5$ is much more  complex, and we are unable to
construct simple and unified formulas.  In the Appendix we list a few
formulas for the lowest $I$'s.

We next point out that  
for  fermions  with  $I \ge (n-2)J - \frac{1}{2} (n-1)(n-2)$
    or  for   bosons with $I \ge (n - 2)J$, 
  $D(I)_J$ is identical for the same $n$ and ${\cal M}$ 
but different $J$.  This identity  is universal and
 exists for all three cases discussed in this paper: 
 fermions in a single-$j$ shell or a single-$l$
 shell, and bosons with  spin $L$. 
Taking $n=5$ as an example, we use
$I=I_{max} - {\cal M}$ to get 
  ${\cal D}({\cal M})_J = D(I)_J$    =1, 0, 1, 1,  2, 2, 3, 3, 5, 5, 
7, 7, 10, 10, 13, 14, 17, 18, 22, 23, 28, 29, 34,
36, 42, 44, 50, 53, 60, 63, $\cdots$, for ${\cal M}=$0, 1, 2,
$\cdots$, 28 $\cdots$, which is independent of $J$  and is 
applicable both to fermions or bosons.  Below we prove 
 this observation. 

According to the standard enumeration procedure \cite{Lawson},
 suppose that
there are $p$ distinct Slater determinants
$\Psi_M(i)$ $(i=1, \cdots p$) with the property $J_z \Psi_M(i) = M
\Psi_M(i)$. If there are $p+q$ ($q > 0$) linearly independent
Slater determinants that have the property
$J_z \Psi_{M-1}(i) = (M-1) \Psi_{M-1}(i)$, then the $q$ states with angular
momentum $I=M-1$ can be constructed.
We use the convention that $J \ge m_1 > m_2  > \cdots > m_n
\ge -J$ for fermions and  
$L \ge m_1 \ge m_2 \ge \cdots \ge m_n \ge -L$ for bosons. 
Here the sum  of $m_i$  over $i$ is equal to 
$M$ for $\Psi_M(i)$ or $(M-1)$ for $\Psi_{M-1}(i)$. One sees that
the number of states of different systems 
is the same if the hierarchies  of $m_i$ for these systems  
have a one-to-one correspondence.

For fermions in a single-$j$ or a single-$l$ shell, 
the hierarchical structure of number of states  
obtained by successively
operating $J_{-}$  on $\Psi_{M_{max}}$ $(m_1, m_2,
\cdots$ $m_n$=$J, J-1, \cdots$, $J-n+1$)  
remains to be  identical  for all $J$'s  until we have
 the case in which  only $m_n$ is decreased to become 
$-J$   while other $m_i$ are unchanged.  In other 
words, before the process of the $J_{-}$ operation is 
 iterated   $2J+2 - n$ times (Here $J=j$ or $l$),
the hierarchy is the same for {\it all} $\Psi_{M}$'s
obtained by  operating $J_{-}$ on   $\Psi_{M_{max}}$ successively: 
If there were not such a  requirement that
$J \ge m_1 > m_2  > \cdots > m_n \ge -J$, the above 
hierarchical structure for fermions would be always the same 
for all $I$ states. 
The minimum   angular momentum $I$ which keeps the
identical hierarchy for $n$ fermions in a single-$J$ shell
is thus given by:
\begin{equation}
(J) + (J-1) + (J-2) + \cdots  (J-n+2) + (-J) =
(n-2)J -
\frac{1}{2} (n-1)(n-2), 
\end{equation}
where $J=j$ or $l$. 

Similarly, for bosons with  spin $L$,
one keeps the hierarchical structure of number of states
while one operates $J_{-}$ on $\Psi_{M_{max}}$
(where $m_1 = m_2 = \cdots = m_n = L$) successively
until one comes to the case in which only $m_n$ is $-L$ and others
unchanged. Thus the minimum  angular momentum $I$ which
keeps the identical hierarchy for $n$ bosons with  spin $L$ is given by:
\begin{equation}
L + L + \cdots +L -L = (n-2)L.
\end{equation}

It is noted  that the  Pauli blocking 
produces a smaller  $I_{max}$ which
equals to $nJ-$  $\frac{1}{2}n(n-1)$  while the $I_{max}$ 
of bosons is $nJ$. Despite of this difference,
there is a one-to-one correspondence between $|m_1, m_2 \cdots m_n \rangle$ 
of bosons and that of fermions in the process of successive operation
of $J_{-}$ on $\Psi_{M_{max}}$. Thus 
$D({\cal M})_J$ for both fermions in a single-$j$ or 
a single-$l$ shell and that of bosons with  spin $L$ are also the same
for the same $n$  but different $J$, 
if $I \ge (n-2)J - \frac{1}{2} (n-1)(n-2)$ for fermions and 
$I \ge (n-2)L$ for bosons.

We  finally address the above discussion  
in  an alternative procedure: 
One  lets $I= I_{max} - {\cal M}$, and uses  ${\cal P}({\cal M})$ to label 
the number of partitions of ${\cal M}=i_1+ i_2 + \cdots  + i_n$ with 
$i_1 \ge i_2   \ge \cdots \ge 0$. 
  Defining ${\cal P}(0)=1$,
one easily finds that $D(I)_J= {\cal D}({\cal M})_J
={\cal P}({\cal M}) - {\cal P}({\cal M}-1)$, unless 
$I < (n-2)J - \frac{1}{2} (n-1) (n-2)$ for fermions
or $I < (n-2)J$ for bosons. This procedure not only
proves  the identity discussed above but also reveals a relevant 
fact: when ${\cal M} \le $min($n$, $2J+1 - n$)  for fermions or 
 ${\cal M} \le $min($n$, $2J$)  for bosons, 
$D(I)_J =  {\cal D}({\cal M})_J $ is the 
same for fermions and bosons, 
because the hierarchy of  ${\cal P}(i)$ of the systems  is identical
for different $n$ and $J$.    For example,
${\cal D}({\cal M}=5)_{J=j=31/2}$  of 5 fermions
is equal to  ${\cal D}({\cal M}=5)_{l=30} = 2 $ of 10 bosons.
The ${\cal D}({\cal M})_{J}$ series are: 1, 0, 1, 1, 2, 2, 4, 4, 7, 8,
12, 14, 21, 24, 34, 41, 55, 66, 88, 105, 137, 165, 210,
253, 320, for ${\cal M}=0, 1, \cdots$ 24, respectively.

To summarize, we  found in this paper
 that there are simple structures in the 
 number of states  of three and four identical particles,
 which enabled us   to construct empirical formulas for $n=3$ and 4. 
   For $n=5$ we presented formulas for a few lowest $I$ states. 
 We have confirmed the validity  of these empirical formulas 
 throughout $j \le 999/2$ and $l\le 500$ which 
 are large enough for practical use. 
                                        
We also  proved  that 
the hierarchy of states obtained by
operating $J_{-}$ on $\Psi_{M_{max}}$ for fermions and bosons 
is the same if $I \ge nJ - \frac{1}{2} (n-1) (n-2)$ for fermions 
or $I \ge (n-2)J$ for bosons, which
gives identical   series of number of states for 
 fermions in a
single-$j$ shell or bosons with  spin $l$  with 
 the same particle numbers.
The number of states is also identical 
for the same ${\cal M}$ but different $n$ and $J$ if
${\cal M} \le $min($n$, $2J+1 - n$)  for fermions and
for ${\cal M} \le $min($n$, $2J$)  for bosons.
These facts have eluded from observation in the long history of 
the enumeration procedure  for the  dimension $D(I)_J$. 

The authors would like to thank Dr. J. N. Ginocchio for his
suggestive discussions on this work and for his reading 
of the manuscript. One of the authors (ZYM) is grateful to 
  the Japan Society for the Promotion of Science 
 (contract No. P01021) for supporting his work.

\vspace{0.5in}

{ Table I ~  ~ For the cases with $n=4$,
  $\eta_I^J$, $\eta_I^{J+1}$ and
$\eta_I^{J+2}$ of angular momenta $I$ and $I+3$ ($I$ is even) states 
change periodically at an interval $\Delta_J=3$ when $I \le 2J$.
One thus  easily constructs formulas for  $D(I)_J$'s
for states with $I \le 2J$.  }

\vspace{0.2in}

\begin{tabular}{cc|ccc} \hline \hline
 $I$ & $I+3$ ~  & $~~\eta_I^J$ & $\eta_I^{J+1}$ & $\eta_I^{J+2}$  \\ \hline
0   &   3 ~ & ~~ 1 & 0 & 0   \\
2   &   5 ~ & ~~ 1 & 1 & 0   \\
4   &   7 ~ & ~~ 1 & 1 & 1   \\
6   &   9 ~ & ~~ 2 & 1 & 1   \\
8   &   11 ~ & ~~ 2 & 2 & 1   \\
10  &   13 ~ & ~~ 2 & 2 & 2   \\
12  &   15 ~ & ~~ 3 & 2 & 2   \\
14  &   17 ~ & ~~ 3 & 3 & 2   \\
16  &   19 ~ & ~~ 3 & 3 & 3   \\
$\vdots$  &  $ \vdots$ ~ & ~~ $\vdots$ &  $\vdots$   & $\vdots$ \\ \hline \hline
\end{tabular}

\vspace{0.3in}

\newpage

\begin{center}
Appendix ~~  number of states for five particles lying in a few 
lowest $I$ states.
 
\end{center}

First, we come
to the case with five fermions in a single-$j$ shell. 
We define $A(j, j_0, d) = \left[ \frac{j-j_0}{d}  \right]$, 
and $B(j, j_0, d) = (j-j_0) \% d$, and obtain  
\begin{eqnarray}
&& D(\frac{1}{2})_j = 6 A^2(j, \frac{9}{2}, 12) 
+ 3 A(j, \frac{9}{2}, 12) \nonumber \\
& & + \left(A(j, \frac{9}{2}, 12) + 1 \right)
\left(B(j, \frac{9}{2}, 12) +1 \right) + \delta_{B(j, \frac{9}{2}, 12)} ~, 
\end{eqnarray}
where
\begin{eqnarray}
&& \delta_{B(j, \frac{9}{2}, 12)}   =  \left\{
\begin{array}{ll}
- B(j, \frac{9}{2}, 12)      &  {\rm if}~
B(j, \frac{9}{2}, 12) \le 2 ~     \\
-2      &  {\rm if}~  3 \le B(j, \frac{9}{2}, 12) \le 4 ~   \\
-3      &  {\rm otherwise} ~  
\end{array}  \right.      ~~ ;      \nonumber
\end{eqnarray}
\begin{eqnarray}
&& D(\frac{3}{2})_j = 3 A^2(j, \frac{11}{2}, 6) 
+ 4 A(j, \frac{11}{2}, 6) \nonumber \\
& & + \left( A(j, \frac{11}{2}, 6) + 1 \right)
\left( B(j, \frac{11}{2}, 6) +1 \right) +
\delta_{B(j, \frac{11}{2}, 6)} +1 ~,  \label{not}
\end{eqnarray}
where
\begin{eqnarray}
&& \delta_{B(j, \frac{11}{2}, 6)}   =  \left\{
\begin{array}{ll}
 1      &  {\rm if}~   B(j, \frac{11}{2}, 6) \ge 4    \\
 0      &  {\rm otherwise}    
\end{array}  \right.     ~ ~ ;     \nonumber
\end{eqnarray}
\begin{eqnarray}
&& D(\frac{5}{2})_j = 2 A^2(j, \frac{9}{2}, 4) 
+ 3 A(j, \frac{9}{2}, 4) \nonumber \\
& & + \left( A(j, \frac{9}{2}, 4) + 1 \right)
\left( B(j, \frac{9}{2}, 4) +1 \right) + \delta_{B(j, \frac{9}{2}, 4)} +1 ~, 
\end{eqnarray}
where
\begin{eqnarray}
&& \delta_{B(j, \frac{9}{2}, 4)}   =  \left\{
\begin{array}{ll}
 1      &  {\rm if}~   B(j, \frac{9}{2}, 4) \ge 2    \\
 0      &  {\rm otherwise}    
\end{array}  \right.    ~ ~ ;       \nonumber
\end{eqnarray}
and 
\begin{equation}
  D(\frac{7}{2})_j = 6 A^2(j, \frac{3}{2}, 6) 
  + \left( 2  A(j, \frac{3}{2}, 6) + 1 \right)
\left( B(j, \frac{3}{2}, 6) +1 \right)
  + \delta_{B(j, \frac{3}{2}, 6)} -1 ~, 
 \label{final}
\end{equation}
where
\begin{eqnarray}
&& \delta_{B(j, \frac{3}{2}, 6)}   =  \left\{
\begin{array}{ll}
-1      &  {\rm if}~   1 \le  B(j, \frac{3}{2}, 6) \le 3   \\
 0      &  {\rm otherwise}   
\end{array}  \right.  ~ ~ .         \nonumber 
\end{eqnarray}

It is noted that $D(\frac{3}{2})_j = 1$ for $n=5$ when $j=\frac{9}{2}$, which
is not covered in the formula (\ref{not}). 
An interesting behavior is that
there exists an approximate relation
for 5 fermions in a single-$j$ shell: 
$D(I)_j \sim (I+\frac{1}{2}) D(\frac{1}{2})_j$ when $I < j$.

Next, we come to five fermions in a single-$l$
shell. For $I=0$, we define
\begin{eqnarray}
&& k   =  \left\{
\begin{array}{ll}
 (l-2)/2       &  {\rm if}~ l\% 2 =0      \\
 (l-11) /2       &  {\rm if } ~l\% 2 =1   
\end{array}  \right.  ~, ~ ~ K = \left[ k/6 \right], ~ ~ 
{\cal K} = k \% 6 ~,  \nonumber 
\end{eqnarray}
and obtain 
\begin{equation}
D(0)_l = 3 K
\left( K+1\right) - K
+({\cal K} +1) \left(K   +1\right) + \delta_{{\cal K} 0} -1 ~ ; 
\end{equation} 
for $I=1$, we define
\begin{eqnarray}
&& k   =  \left\{
\begin{array}{ll}
 (l-1)/2       &  {\rm if}~ L\% 2 =1    \\
 (L-4)/2        &  {\rm if } ~L\% 2 =0  
\end{array}  \right. ~, ~~ K = \left[ k/2 \right], ~~
{\cal K} =
k \% 2 ~ ~,   \nonumber
\end{eqnarray}
and obtain 
\begin{equation}
D(1)_l = K\left( K+1 \right) + 
{\cal K} (K+1 ) ~~.
\end{equation}

We finally come to the case of  five bosons with  spin $L$. 
For $I=0$, we define 
\begin{eqnarray}
&& k   =  \left\{
\begin{array}{ll}
 L/2       &  {\rm if}~ L\% 2 =0      \\
 (L-9) /2       &  {\rm if } ~L\% 2 =1   
\end{array}  \right. ~ ~,    ~~ K = \left[ k /6  \right], ~~
{\cal K} =  k \% 6 ~,   \nonumber
\end{eqnarray}
and obtain 
\begin{equation}
D(0)_L = 3 K
\left( K +1\right) - K
+ ({\cal K}+1)  \left( K +1\right)
+ \delta_{{\cal K}0} -1  ~; 
\end{equation}
for $I=1$, we define
\begin{eqnarray}
&& k   =  \left\{
\begin{array}{ll}
 L/2       &  {\rm if}~ L\% 2 =1      \\
 (L-3)/2        &  {\rm if } ~L\% 2 =0  
\end{array}  \right. ~ ~ , ~~ K =  \left[ k /2  \right], ~~ 
{\cal K} =
 (k  \% 2)  ~ ~,      \nonumber
\end{eqnarray}
and obtain 
\begin{equation}
D(1)_L =  \left( K +1 \right)
(K+ {\cal K}   +1 ) ~~.
\end{equation}

The formulas for larger $I$'s with $n=5$ 
are more complicated and are not 
addressed  in this paper. 

\end{document}